\documentclass[runningheads]{llncs}
\usepackage[T1]{fontenc}
\usepackage{booktabs} 
\usepackage{amsmath,amsfonts}
\usepackage{graphicx}
\usepackage{todonotes}
\usepackage{tikz}
\usepackage{enumitem}
\usepackage{booktabs} 
\usepackage{booktabs}
\usepackage{subcaption}
\usepackage{multicol}
\usepackage[ruled,linesnumbered,longend]{algorithm2e}
\usepackage{xspace}
\usepackage{tabularx}
\usepackage{cite}
\usepackage{subcaption}
\usepackage[firstpage,pos={0.99\paperwidth,0.5in},hanchor=r]{draftwatermark}
\usepackage[
  colorlinks = true,
  linkcolor = blue,
  urlcolor  = blue,
  citecolor = blue,
  anchorcolor = blue]{hyperref}

\begin{document}

\RestyleAlgo{ruled}
\SetAlgoNoEnd
\SetKwComment{Comment}{/* }{ */}

\newcommand{\sysname}{Justin\xspace}
\newcommand{\flink}{Flink\xspace}
\newcommand{\flinklong}{Apache Flink\xspace}
\newcommand{\kafka}{Kafka\xspace}
\newcommand{\kafkalong}{Apache Kafka\xspace}

\newcommand{\mypara}[1]{\smallskip\noindent{\bf {#1}.}}

\newcommand{\omnifiguresscalingfactor}{0.5}
\newcommand{\matplotlibscalingfactor}{0.545}

\ifthenelse{\boolean{false}}
{ \newcommand{\mynote}[3]{
    \fbox{\bfseries\sffamily\scriptsize#1}
    {\small$\triangleright$\textsf{\emph{\color{#3}{#2}}}$\triangleleft$}}}
{ \newcommand{\mynote}[3]{}}
\newcommand{\ds}[1]{\mynote{Donatien}{#1}{magenta}}
\newcommand{\er}[1]{\mynote{Etienne}{#1}{blue}}
\newcommand{\TODO}[1]{\mynote{TODO}{#1}{red}}

\newcommand*\circled[1]{\tikz[baseline=(char.base)]{
            \node[shape=circle,draw,inner sep=1pt] (char) {#1};}}

\newcounter{takeawayCounter}
\setcounter{takeawayCounter}{0}

\newcommand{\takeaway}{
    \refstepcounter{takeawayCounter}
    \smallskip\noindent{\bf \emph{Takeaway \arabic{takeawayCounter}}.}
    }

\newcommand{\vspacebeforesection}{\vspace{-2mm}}
\newcommand{\vspaceaftersection}{\vspace{-2mm}}
\newcommand{\vspacebeforesubsection}{\vspace{-2mm}}
\newcommand{\vspaceaftersubsection}{\vspace{-1mm}}

\newcommand{\vspacebeforecaption}{\vspace{-0mm}}
\newcommand{\vspaceaftercaption}{\vspace{-0mm}}

\SetWatermarkAngle{0}
\SetWatermarkText{\includegraphics[]{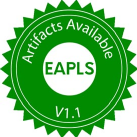}}
\title{\sysname: Fine grain resource management for Distributed Stream Processing}
\title{\sysname: Fine-Grain Memory Allocation for Distributed Stream Processing}
\title{\sysname: Fine-Grain Memory and CPU Scaling for Distributed Stream Processing}
\title{\sysname: Elastic Scaling with Fine-Grain Memory Allocation for Distributed Stream Processing}
\title{\sysname: Hybrid CPU/Memory Elastic Scaling for Distributed Stream Processing\thanks{Artifacts available in \url{https://doi.org/10.5281/zenodo.15209785}}}
\titlerunning{\sysname: Elastic Scaling with Fine-Grain Memory Allocation for DSP}
\titlerunning{Hybrid CPU/Memory Elastic Scaling for Distributed Stream Processing}
%
\author{Donatien Schmitz\inst{1} \and Guillaume Rosinosky\inst{2} \and Etienne Rivière\inst{1}}
%

%
\institute{ICTEAM, UCLouvain, Belgium, \email{\{first.last\}@uclouvain.be} \and
IMT Atlantique, Nantes Université, École Centrale Nantes, CNRS, Inria, LS2N - UMR 6004, France, \email{guillaume.rosinosky@inria.fr}}
\maketitle              
\begin{abstract}
    Distributed Stream Processing (DSP) engines analyze continuous data via queries expressed as a graph of operators.
Auto-scalers adjust the number of parallel instances of these operators to support a target rate.
Current auto-scalers couple CPU and memory scaling, allocating resources as one-size-fits-all packages.
This contrasts with operators' high diversity of requirements. 

We present \sysname, an auto-scaler that enables hybrid CPU and memory scaling of DSP operators.
\sysname monitors both CPU usage and the performance of operators' storage operations.
Its mechanisms enable fine-grain memory allocation for tasks upon a query reconfiguration.
The \sysname policy identifies individual operators' memory pressure and decides between adjusting parallelism and/or memory assignment.
We implement \sysname in Apache Flink, extending the Flink Kubernetes Operator and the DS2 CPU-only auto-scaler.
Using the Nexmark benchmark, our evaluation shows that \sysname identifies suitable resource allocation in as many or fewer reconfiguration steps as DS2 and supports a target rate with significantly fewer CPU and memory resources.
%

\end{abstract}
 
\vspacebeforesection
\vspace{-5mm}
\section{Introduction}
\vspaceaftersection
\label{sec:intro}

Stream Processing allows continuous data analysis in a large variety of applications~\cite{fragkoulis2024survey}.
Supporting stream processing over large volumes of incoming data requires distributing computation over multiple machines.
Distributed Stream Processing (DSP) engines, such as Apache Storm~\cite{storm}, Spark Streaming~\cite{zaharia2013discretized}, or Apache Flink~\cite{flink}, emerged to address the many challenges associated with the distribution and orchestration of parallel stream processing.
This includes fault tolerance~\cite{carbone2017state}, connection to input sources and destinations, and support for elastic scaling, i.e., the ability to dynamically adapt the amount of computational resources to sustainably support a target \emph{rate} of input events.

A stream processing query is a directed graph of \emph{operators} performing computation on incoming events.
Elastic scaling assigns, at runtime, each operator to several \emph{tasks}, i.e., individual threads of execution.
The flow of events is distributed to these tasks to be processed in parallel.
Numerous auto-scalers have been proposed in the literature~\cite{roger2019comprehensive,cardellini2022runtime}.
\flinklong recently integrated a variant of the DS2~\cite{kalavri2018three} auto-scaler in its Kubernetes Operator~\cite{k8sflink}.
DS2 determines the processing capacity of tasks based on a measure of their \emph{busyness}, the fraction of CPU time effectively spent processing events.
It derives a new configuration mapping operators to a required number of tasks to sustain a target input rate while considering cascade effects between operators' loads.

\mypara{Motivation}
Existing auto-scalers couple CPU and memory allocation~\cite{carbone2017state}.
Scaling out an operator (i.e., adding more tasks) adds a proportional amount of memory.
This coupled, one-size-fits-all allocation clashes with the heterogeneous memory requirements of operators.
Some operators, such as a \emph{filter} or a \emph{map}, are stateless and do not require more than a minimal amount of memory to process events efficiently.
In contrast, other operators, such as \emph{joins} or \emph{group by} over long windows, maintain state across the processing of many events~\cite{verwiebe2023survey}.

Intuitively, stateful operators with under-allocating memory perform poorly: The necessary state may not fit in the main memory, and accesses may resort to on-disk storage.
In contrast, ample memory allocation (e.g., as the operator has been scaled out to multiple tasks) should not yield better performance than a smaller but sufficient allocation.

In reality, the relation between event processing performance, state access latency, and memory requirements is more subtle than this simple intuition.
State-of-the-art state backends for stream processing use a Log-Structured-Merge tree (LSM)~\cite{o1996log,luo2020lsm}.
An LSM combines tables and caches in memory with bulk storage on disk.
They use optimizations to favor write performance and minimize the wear of modern SSD drives.
For instance, \flink uses the LSM-based RocksDB~\cite{dong2021rocksdb,rocksdb} for production deployments.
The performance in response to the available memory of RocksDB depends heavily on the nature of the workload~\cite{asyabi2022new}.
Workloads formed mainly of \emph{write} operations do not benefit from a large memory allocation.
In contrast, workloads dependent on \emph{read} operations do, in proportion to the size of their working set.
The nature of an operator's access to its state is not predictable, as it often depends on the characteristics of its input events.

\mypara{Contributions}
We present \sysname, a hybrid auto-scaler that considers \emph{both} CPU and memory allocation when reconfiguring DSP queries.
In contrast with previous approaches, \sysname's auto-scaling policy can decide to scale \emph{up} (i.e., adding more memory to each task) or scale \emph{out} (adding more tasks) depending on runtime indicators on state access performance.
We integrate \sysname in \flinklong and the Flink Kubernetes Operator~\cite{k8sflink}, extending the current elastic scaling engine based on DS2~\cite{kalavri2018three}.

Our contributions and the outline of the remainder of this paper are as follows.
We start with the necessary background and terminology about \flink (\S\ref{sec:background}).
We discuss state management and RocksDB and analyze, using microbenchmarks, the impact of workload characteristics on state lookup performance and response to memory availability (\S\ref{sec:motivations}).

We then present \sysname policies and mechanisms (\S\ref{sec:justin}).
We detail DS2, \flink's current elastic scalers.
We present how \sysname collects metrics about a running query's resource usage and storage performance in RocksDB.
Based on these metrics, \sysname's auto-scaling policy adjusts DS2 decisions and derives CPU and memory allocations.
These allocations are enacted by mechanisms for fine-grain resource allocation in \flink and the Flink Kubernetes Operator~\cite{k8sflink}.

We evaluate \sysname in \flink on a 7-node cluster under high loads and compare it to DS2 using the Nexmark benchmark~\cite{tucker2008nexmark} (\S\ref{sec:evaluation}).
Our results show \sysname produces configurations with 48\% less CPU and 27-28\% less memory for \texttt{q8} and \texttt{q11}, two complex stateful queries while requiring the same or fewer reconfiguration steps than DS2.

We review related work on hybrid vertical/horizontal scaling (\S\ref{sec:related}) and conclude the paper by identifying future directions (\S\ref{sec:conclusion}).

\vspacebeforesection
\section{Background}
\vspaceaftersection
\label{sec:background}

\begin{figure}[t]
    \centering
    \includegraphics[scale=\omnifiguresscalingfactor]{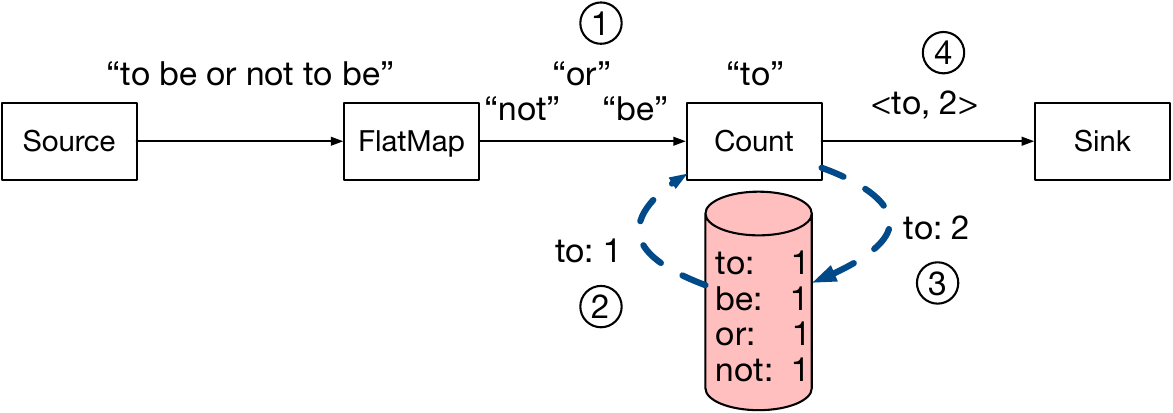}
    \vspace{-2mm}
    \caption{A WordCount query using four operators.\vspace{-5mm}}
    \label{fig:wordcount}
\end{figure}

We provide background information on stream processing.
We use \flink's terminology but note that concepts are similar in other stream processing engines.
We detail state management and elastic scaling in Sections~\ref{sec:motivations} and~\ref{sec:justin}.

\mypara{Stream processing}
A stream processing query is a dataflow graph where vertices are \emph{operators} $o_i \in O$ and edges represent flows of events between these operators.
Operators can support arbitrary computation but are often selected from a library of classical transformations or compiled from SQL~\cite{rabl2016apache}.

Operators can be single-input, such as a \emph{filter} (i.e., letting through a subset of events), a \emph{map} (individual transformation of events), or a \emph{group by} followed by aggregation. 
Others consume multiple streams, e.g., \emph{joins}.
Commonly, operators such as aggregates and \emph{joins} compute over a \emph{window} defined as a count of events or a period~\cite{verwiebe2023survey}.
These operators are \emph{stateful}: they must keep track of information related to events over this window.
In contrast, operators that process events in isolation (\emph{map}, \emph{filter}, etc.) are \emph{stateless}.

\mypara{Word count example}
Figure~\ref{fig:wordcount} presents \emph{word count}, a classic example query implemented using three operators.
This query counts occurrences of individual words in a stream of sentences.
The \emph{Source} is a specific operator injecting sentences as new events, e.g., from disk or a stream store such as \kafkalong~\cite{kreps2011kafka}.
A \emph{FlatMap} operator splits these sentences into multiple events, one for each word (\circled{1}).
The \emph{Count} operator combines a \emph{group by} and a \emph{sum} aggregate to count occurrences of each word over a time window.
The processing of each event by this operator requires fetching the current count from the storage backend (\circled{2}) and updating it (\circled{3}).
Finally, a \emph{Sink} operator receives and stores outputs, e.g., to \kafka (\circled{4}).
In this example, the \emph{Count} operator is stateful; others are stateless.

\begin{figure}[t]
    \centering
    \includegraphics[scale=\omnifiguresscalingfactor]{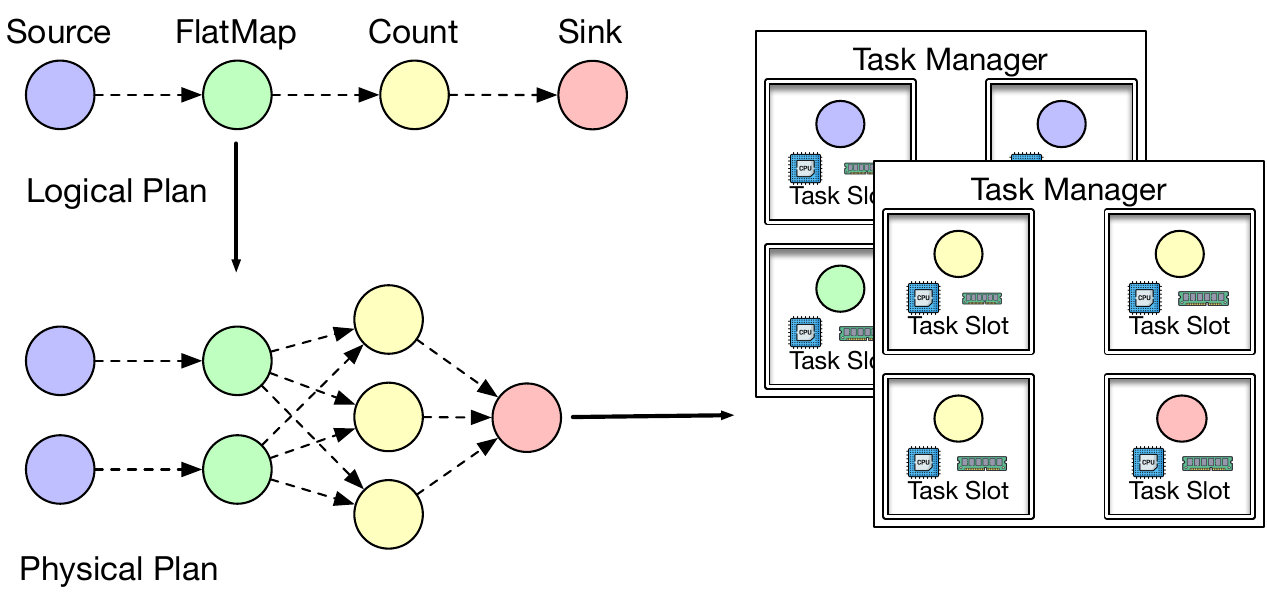}
    \vspace{-2mm}
    \caption{
    \label{fig:flink}  
    Word count query deployed on two Task Managers and eight Task Slots.\vspace{-5mm}
    }
\end{figure}

\mypara{Query execution}
A query's logical plan is executed by a physical plan where each operator can be supported by several processing \emph{tasks}, as illustrated by Figure~\ref{fig:flink}.
Each task is a single thread of execution.
The number of tasks of an operator is also denoted as its \emph{parallelism}.
In \flink, the execution of tasks is supported by a fleet of processes (Java Virtual Machines) called Task Managers (TMs) orchestrated by a centralized Job Manager (JM).
Each TM has a pre-defined number of identical Task Slots (TSs).
Each TS can run one task.
CPU resources allocated to a TM are divided between its TSs.
In this paper, we consider the standard one-core-per-task model.
We discuss the allocation of TM memory in the next section and evaluate its impact on performance.


\vspacebeforesection
\section{Impact of Memory on Operators' Performance} 
\vspaceaftersection
\label{sec:motivations}

We study in this section the relationship between memory allocation and task performance in \flink.
The takeaways of this section guide the design of our hybrid CPU/memory auto-scaler, \sysname, detailed in the next section.


\begin{figure}[t]
    \centering
    \includegraphics[scale=\omnifiguresscalingfactor]{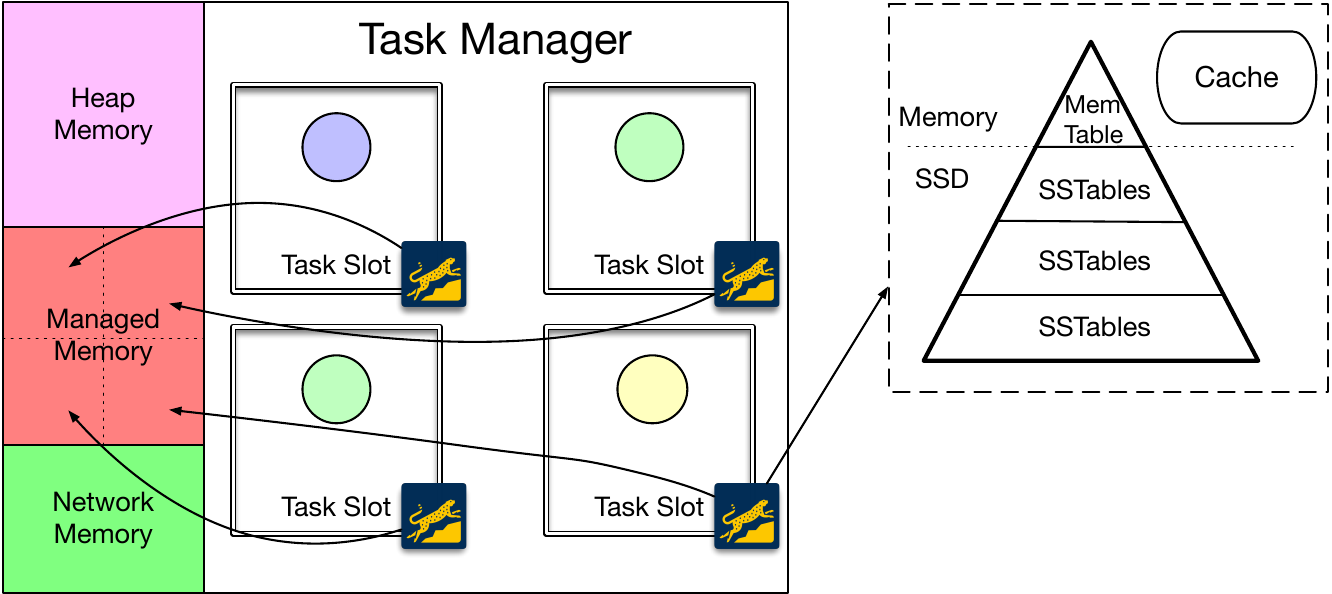}
    \vspace{-2mm}
    \caption{Memory allocation between tasks and RocksDB LSM tree.\vspace{-5mm}}
    \label{fig:flink-memory}  
\end{figure}

\mypara{RocksDB}
The state of stateful operators' tasks must be persisted across the processing of different events.
The storage must allow for reconfiguration and restoration after failures~\cite{carbone2017state}.
Furthermore, this state may be arbitrarily large and not fit into memory.
While an in-memory storage backend is available for testing in \flink, production deployments use RocksDB~\cite{dong2021rocksdb,rocksdb}, a state backend using a Log Structured Merge (LSM) tree and offering a key/value interface.

An LSM tree, illustrated on the right of Figure~\ref{fig:flink-memory}, is a data structure optimized for write-heavy workloads and reducing wear on the storage device, typically an SSD.
Writes are buffered in memory (in a MemTable) and later flushed to disk in sorted order, reducing random I/O.
A hierarchy of sorted SSTables (Sorted String Tables) files is periodically merged into larger ones, reducing fragmentation and read amplification.
Data is stored across multiple levels of this hierarchy, with newer data at higher levels and older data compacted into lower levels.
As a result, reads may require searching SSTables across multiple levels, increasing their latency.
To mask these costs, RocksDB uses an in-memory \emph{cache} of recently accessed data. 

The MemTable, implemented as a skip list, is used to buffer writes before consolidating them to SSTables.
Using a large MemTable has no real impact on write latencies, as consolidation is performed at the granularity of the first-level SSTable (of 64~MB).
In contrast, read latency is directly impacted by the size of the cache and its relation to the task's working set size (i.e., the set of keys the task frequently accesses over a period).
Tasks performing frequent reads, as for the \emph{Count} operator of the word count in Figure~\ref{fig:wordcount}, may have to resort to costly exploration of SSTables on disk if the cache is not large enough.
Tasks performing only or mostly writes are not impacted by the cache size.

\mypara{TM memory allocation}
Memory sharing between TSs on the same TM depends on the JVM's memory segmentation.
On-heap memory, used by \flink operators for creating Java objects and subject to garbage collection, is shared among all tasks.
The same applies to network memory used for communication buffers.
While there is no isolation between threads for access to these segments, a TM reserves a minimum amount for each TS.
In contrast, \emph{managed memory} is specific to a thread and is not subject to garbage collection.
It is used by the RocksDB instance local to each task to store its MemTable and cache. 

\takeaway
Managed memory is reserved for all task slots in equal amounts.
This memory is wasted for stateless tasks that do not use RocksDB.
Stateful tasks get a one-size-fits-all allocation, regardless of their requirements.

\begin{figure}[t]
    \centering
    \includegraphics[scale=\matplotlibscalingfactor]{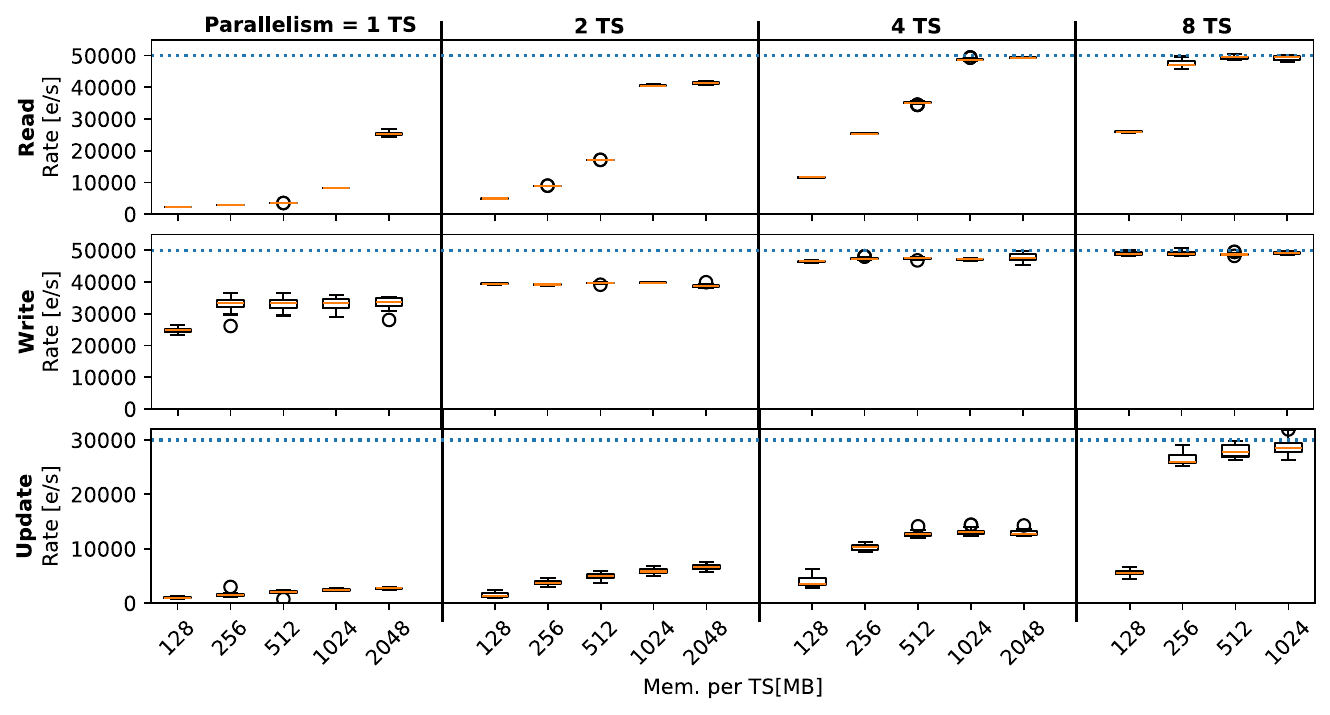}
    \vspace{-7mm}
    \caption{Evaluation of multiple memory-cpu configurations on three different state access patterns.
    Benchmarks show the maximum rate reachable against a target rate (dotted line).
    The impact of memory on rate significantly differs depending on the workload, read-only being the more profitable.\vspace{-5mm}
    }
    \label{fig:microbenchmarks}  
\end{figure}


\mypara{Microbenchmarks}
We evaluate the impact of memory allocation on performance using microbenchmarks.
We use a single operator.
The input stream is formed of events of 1,000~B.
Each event includes a key generated uniformly at random between 0 and 1,000,000 and a random payload.
The RocksDB state backend is pre-populated with a value associated with each key.
We consider three workloads.
In the \textbf{Read} workload, the operator reads the value associated with the key in the event.
In the \textbf{Write} workload, it replaces the value associated with this key without reading the previous value.
In the \textbf{Update} workload, the operator reads the current value and then overrides it with the event's payload.
We use a target rate of 50,000 events per second for the Read and Write workloads and 30,000 events per second for the Update workload.

Our results are presented in Figure~\ref{fig:microbenchmarks}.
We consider 19 configurations for each workload, with an operator's parallelism ranging from 1 to 8 tasks and managed memory allocation of 128 to 2,048~MB.
In any memory allocation, the MemTable size is at most 64 MB, and the rest is used for the cache.
By default, \flink prioritizes the allocation of at least half of the memory to the cache, possibly reducing the size of the MemTable, which is allocated as a power-of-2 granularity. 
As a result, an allocation of 128~MB of memory results in a 32~MB MemTable and a 96~MB cache, but allocations of 256~MB or 512~MB result in a 64~MB Memtable and, respectively, 192~MB and 448~MB caches.
We denote a configuration with $t$ tasks and $m$~MB of memory as ($t;m$).
We run each configuration for 10 minutes.
We measure aggregate rate every 5 seconds and present the distribution as a box plot.
The dotted line indicates the target rate.
We use specific source operators that produce events at the maximal possible speed, subject to back pressure from the measured operator and capped by this target rate.

We observe that the target rate is sustained for the Read workload starting from (4; 1,024) or (8; 512).
Below that, we observe that the cache hit rate (not shown on the plot) is very low, impacting processing time for every event.
With 256~MB of memory, even the 8-task configuration is slightly below the target rate, possibly requiring further scale-out and the use of a lot of CPU resources.

\takeaway Scaling out a read-intensive operator without appropriate memory allocation can lead to inefficient scale-out operations.
The cache hit rate is a key metric for determining when the cache size is insufficient and scaling up.

The Write workload shows constant performance at all parallelisms with all memory configurations, except the smallest (1; 128), which is slightly below (1; 256).
This lower performance is due to the smaller MemTable size.
The target rate is reached with a parallelism of 8, with 4 tasks being very close.

\takeaway The size of the cache does not impact write performance.
Write-dominated operators should favor scale-out to scale-in.

Finally, the evaluation of the Update workload shows that only an 8-task configuration can sustain the workload.
We also observe a \emph{plateau} effect for 4 and 8 tasks, where adding memory does not result in significant gains (in contrast with the Read workload, where these gains are more linear).
Configurations with insufficient memory (128~MB) cannot sustain the rate regardless of parallelism.

\takeaway It is necessary to vertically scale (add memory) above a minimum threshold.
If a scale-up does not improve performance significantly, it is preferable to scale out.

\vspacebeforesection
\section{\sysname: Hybrid CPU/memory Auto-Scaler}
\vspaceaftersection
\label{sec:justin}

Building upon the takeaways of the previous section, we design \sysname as a hybrid auto-scaler that decides on scaling \emph{up} or scaling \emph{out} an operator depending on its needs and avoids allocating memory to tasks that do not need or use it.
\sysname builds upon DS2~\cite{kalavri2018three}, \flink's current auto-scaler.
We start with a recap of DS2 principles.
Then, we detail the metrics collected by \sysname.
We present an auto-scaling policy that uses these metrics and explain how we enable dynamic, heterogeneous memory allocation for tasks.

\mypara{Elastic scaling and DS2}
Elastic scaling determines the appropriate level of parallelism for all operators.
From a default configuration (parallelism of tasks and memory per task) at time $t=0$, reconfigurations are triggered, based on observations, at discrete times $t>0$.
We define a configuration $C^t$ as a map between operators $o_i \in O$ where $o_i.p^t$ is the parallelism $p$ (number of tasks) of operator $i$ at time $t$.

DS2~\cite{kalavri2018three} uses a primary metric, the busyness of operators, averaged across their tasks.
Busyness is the fraction of CPU time spent processing events, i.e., not idling or waiting due to back pressure.
A reconfiguration trigger is a high busyness for one of its operators in addition to backpressure from its upstream operator(s), indicating that the query's capacity is insufficient to cope with the current rate. 
When triggered at time $t$, DS2 determines a new level of parallelism $o_i.p^t$ for each operator such that the resulting busy rate is below a target.
For this purpose, it uses linearity assumptions and considers that the load will be equally distributed among the scaled-out operator's tasks.
This computation accounts for the cascade effects between operators: scaling out an operator increases its capacity, resulting in a higher rate for downstream operators and the need to scale them out as well.
The new configuration is applied via a query reconfiguration, possibly resulting in state transfers between RocksDB instances~\cite{carbone2017state}.
DS2 typically requires several reconfiguration steps until it reaches a stable configuration for a target rate, and setting appropriate thresholds for triggering reconfiguration and target busyness levels requires careful tuning.

\mypara{Integrating memory-awareness}
%
\sysname requires memory-centric indicators in addition to CPU ones, such as busyness.
We note, however, that there can be a coupling between memory and CPU indicators.
A task performing high-latency state accesses may have a high busyness, as state accesses are accounted for as part of the event processing time.
State access latency allows distinguishing between the two cases. 
High average access latency $o_i.\tau^t$ indicates that a significant fraction of accesses by $o_i$'s tasks used the disk and that scaling out may not be the best option until the cache is appropriately dimensioned.
Furthermore, RocksDB exports the number of cache hits and misses, from which we can derive an average \emph{cache hit rate} $o_i.\theta^t$ for operator $o_i$'s tasks.
We collect these two metrics using Prometheus~\cite{prometheus}.
As for other metrics, these are collected and averaged over a period.


\vspacebeforesubsection
\subsection{\sysname Policy Building BLocks}
\vspaceaftersubsection

The elastic scaling policy of \sysname extends DS2, specifically its implementation in \flink and the associated Kubernetes support operator~\cite{k8sflink}.
This choice allows building upon the complex engineering, integration, and parameter tuning already made by the open-source community and improves impact potential.

\mypara{Memory allocation}
\sysname allocates heterogeneous memory to the tasks of different operators.
$o_i.m^t$ represents in a configuration $C^t$ the managed memory, in MB, allocated to all tasks of an operator $o_i$.
To facilitate the mapping of tasks to task managers, we allocate memory in \emph{levels}.
Each new level corresponds to double per-task memory, from a minimum to a maximum (\emph{maxLevel}).
For $o_i.m^t = x$, we allocate $2^{x}$ times the minimum of managed memory for stateful tasks.
If, for instance, the minimum managed memory is 128~MB, $o_i.m^t=0$ means 128~MB, $o_i.m^t=1$ means 256~MB, etc.
Stateless tasks are allocated no managed memory, expressed as $o_i.m^t=\bot$.

\mypara{Decisions history}
DS2 does not maintain a history of scaling decisions, as it only takes these in one direction (horizontally).
In contrast, \sysname can choose between scaling out or up.
Keeping a decision history helps determine whether they improved the query capacity.
Past configurations are available in $C^0 \dots C^{t-1}$.
A boolean $o_i.v^t$ indicates that, in $C^t$, the scaling decision involved \emph{scaling up} (vertically) the memory of operator $o_i$'s tasks (i.e., $o_i.v^t = \top \implies o_i.m^t > o_i.m^{t-1}$).

\SetKwInput{KwData}{Parameters}
\SetKwComment{Comment}{// }{}
\newcommand\mycommfont[1]{\footnotesize\ttfamily\textcolor{purple}{#1}}
\SetCommentSty{mycommfont}

\SetKwIF{If}{ElseIf}{Else}{if}{}{else if}{else}{end if}%

\begin{algorithm}[t!]
    \caption{\sysname's hybrid elastic scaling policy}\label{alg:policy}
    \DontPrintSemicolon
    \KwData{$\Delta_\theta  \gets \text{\emph{80\%}}$; $\Delta_\tau  \gets 1\text{\emph{ms}}$; $\text{\emph{maxLevel}} \gets 3$}
    \KwResult{A new configuration $C^t$}
    
    \BlankLine
    
    $C^t \gets \text{DS2()}$ \label{alg:policy:getds2} \Comment*[r]{Obtain initial configuration from DS2}
    \For(\tcp*[f]{Iterate over all operators}){$o_i \in C^t$}{
      \eIf(\tcp*[f]{No recorded RocksDB access?}){$o_i$ is stateless}{ \label{alg:policy:is_stateless}
        $o_i.m^t \gets \bot$ \Comment*[r]{Disable managed memory for $o_i$} \label{alg:policy:stateless_remove_mem}
      }{\If(\tcp*[f]{$o_i$'s capacity insufficient?}){$o_i.p^t \neq o_i.p^{t-1}$ \label{alg:policy:capacity-insufficient}} 
      {\eIf(\tcp*[f]{Vertical scaling used last time?}){$o_i.v^{t-1}$\label{alg:policy:previous_vertical}}{
        \eIf(\tcp*[f]{Did it improve?}){$(o_i.\theta^t > o_i.\theta^{t-1} \vee o_i.\tau^t < o_i.\tau^{t-1})$ \label{alg:policy:vertical_improved}}{ \label{alg:policy:further_scaling_condition} 
        \If(\tcp*[f]{Can scale-up?}){$(o_i.m^{t-1} + 1) <$ maxLevel \label{alg:policy:vertical_improved_canscaleup}}{
          $o_i.p^t\gets o_i.p^{t-1}$ \Comment*[r]{Cancel scale out} \label{alg:policy:cancel_1}
          $o_i.m^t\gets o_i.m^{t-1} + 1$ \Comment*[r]{Increase memory further}\label{alg:policy:increase_1}
          $o_i.v^t \gets \text{True}$\;
        }
        }(\tcp*[f]{Did not improve?}){
            $o_i.m^t\gets o_i.m^{t-1} - 1$ \Comment*[r]{Roll-back scale in}\label{alg:policy:roll_back_1}
        }
            
    }{
    
    \If(\tcp*[f]{Could vertical scaling be useful?}){$(o_i.\theta^t < \Delta_\theta \vee o_i.\tau^t > \Delta_\tau) \wedge o_i.m^{t-1} + 1 < $ maxLevel \label{alg:policy:below_thresholds}}{ \label{alg:policy:new_scaling_condition}
        $o_i.p^t\gets o_i.p^{t-1}$ \Comment*[r]{Cancel scale out}\label{alg:policy:cancel_2}
        $o_i.m^t\gets o_i.m^{t-1} + 1$ \Comment*[r]{Attempt increasing memory}\label{alg:policy:increase_2}
        $o_i.v^t \gets \text{True}$ \label{alg:policy:end} \;
    }
    }
        
    }}
    }
    \Return $C^t$ \Comment*[r]{Return the updated configuration}
\end{algorithm}


\vspacebeforesubsection
\subsection{\sysname Policy Algorithm}
\vspaceaftersubsection

Algorithm~\ref{alg:policy} is called when the capacity of the query is insufficient and requires a reconfiguration.
We use the unmodified DS2 trigger based on busyness and backpressure metrics.




The key principle of \sysname's policy is, in appropriate situations, to prioritize a vertical scaling action to a horizontal scaling action decided by DS2.
It uses memory efficiency metrics to determine if there is a gain potential and leverages insights from past scaling decisions to evaluate whether they improved capacity.
The policy starts by calling the unmodified DS2 (Line~\ref{alg:policy:getds2}).
It then iterates over all operators.
It first identifies if the operator is stateless, which is indicated by the absence of RocksDB-specific metrics in Prometheus (Line~\ref{alg:policy:is_stateless}).
If it is, we remove its portion of managed memory (Line~\ref{alg:policy:stateless_remove_mem}) and keep the parallelism given by DS2.

For stateful operators, we consider those where a re-scaling decision is proposed by DS2 (Line~\ref{alg:policy:capacity-insufficient}).
Other operators' capacities are deemed sufficient and do not need to scale.
We then distinguish between two cases: if the operator was vertically scaled the previous time (Lines~\ref{alg:policy:previous_vertical}--\ref{alg:policy:roll_back_1}) or not (Lines~\ref{alg:policy:below_thresholds}--\ref{alg:policy:end}).

If the operator was previously scaled up, we determine if this led to an improvement.
As discussed in Section~\ref{sec:motivations}, this depends on the number of writes vs. reads, the size of the working set, among other factors.
A scaled-up operator that does not give better capacity is unlikely to improve with further scale-up.
We assess this improvement by comparing the cache hit ratio and state access latency of the current and previous periods (Line~\ref{alg:policy:vertical_improved}).\footnote{The algorithm uses strict inequality signs $<$ and $>$ for clarity, but we can use a minimum amount of improvement as a percentage as well, implementing a hysteresis.}
If there was an improvement and the operator is not already at the maximum memory level, we cancel the scale-out (Line~\ref{alg:policy:cancel_1}) and replace it with scale-up (Line~\ref{alg:policy:increase_1}). 
If there was no improvement, we cancel the previous scale-up (Line~\ref{alg:policy:roll_back_1}) to avoid wasting memory.
The parallelism recommended by DS2 will thus apply using the previous memory configuration.

If an operator was not scaled up already (or was so earlier than $t-1$), we evaluate if the cache hit rate is \emph{below} a threshold $\Delta_\theta$, indicating insufficient cache size for reads or if average state access latency is \emph{over} a threshold $\Delta_\tau$, indicating a significant fraction of operations need costly disk/SSD accesses.
In this case, we cancel the scale-out decision (Line~\ref{alg:policy:cancel_2}) and increase the operator's memory level (Line~\ref{alg:policy:increase_2}).
Otherwise, we apply the recommended parallelism. 

\er{
Notation:
\begin{itemize}
    \item Configurations of the query evolve in discrete time steps $t \geq 0$. 
          At time $t=0$, the query starts with a default configuration.
          The first auto-scaling decision is made at time $t=1$.
    \item A configuration $C^t$ is a map between operators $o_i \in O$ where:
    \begin{itemize}
        \item $o_i.p^t$ is the parallelism $p$ (number of tasks) of operator $i$ at time $t$ (the parallelism in the previous configuration for $t \geq 1$ is $o_i.p^{t-1}$);
        \item $o_i.m^t$ is the managed memory configuration of operator $i$'s tasks at time $t$:
        \begin{itemize}
            \item If $o_i.m^t = \bot$, the operator is considered stateless and is not assigned any managed memory;
            \item otherwise, $o_i.m^t = [ 0, \mathit{max\_level} ]$ is the \emph{memory level} of the operator. A memory level of $o_i.m^t = x$ indicates that $2^{x}$ \emph{slices} of managed memory are allocated to each task of the operator (if a slice is 128~MB, operator $i$'s tasks each get 128~MB at time $t$ if $o_i.m^t = 0$, 256~MB if $o_i.m^t = 1$, 512~MB if $o_i.m^t = 2$, etc.).
        \end{itemize}
        \item $o_i.v^t$ is a boolean indicating if in $C^t$ the scaling decision involved \emph{scaling up} (vertically) the memory of operator $i$'s tasks (i.e., $o_i.v^t = \top \implies o_i.m^t > o_i.m^{t-1}$);
        \item $o_i.h^t$ is a boolean indicating if in $C^t$ the scaling decision involved \emph{scaling out} (horizontally) operator $i$ by adding more tasks (i.e., $o_i.h^t = \top \implies o_i.p^t > o_i.p^{t-1}$).
        \item $o_i.\theta^t$ represents the average cache hit rate during the last window of metric collection.
        \item $o_i.\tau^t$ represents the average state access latency during the last window of metric collection. 
    \end{itemize}
\end{itemize}
}

\vspacebeforesubsection
\subsection{Implementation}
\vspaceaftersubsection

\sysname is implemented in version 1.18 of \flinklong and extends the Kubernetes Operator~\cite{k8sflink} where DS2~\cite{kalavri2018three} is implemented.
Our changes account for about 1,500~LoC.
In addition to the policy described earlier in this section, \sysname relies on mechanisms allowing the enactment of the decisions: (1) the support of Task Slots (TS) with heterogeneous memory allocations; (2) the placement of tasks to TS across different Tasks Managers (TMs), minimizing fragmentation, and (3) the creating when necessary of newer TMs.

\flink 1.18 supports an API allowing to specify custom resource configurations for operators when submitting a new query (number of cores and quantity of heap, network, and managed memory).
We extend \flink's Adaptive Scheduler module to allow such changes to be enacted at runtime via a REST API.
The scheduler maps heterogeneous memory demand to the available TMs using a standard multidimensional Bin-Packing algorithm~\cite{lodi2002recent}. \er{We could have a bit more detail here.}

The \flink policy is implemented as part of \flink's Kubernetes Operator~\cite{k8sflink}.
An API allows the policy's parameters, such as trigger or decision thresholds, to be dynamically adapted at runtime.
This allows control over the elastic scaling of long-running queries without redeployment.
When the Bin-Packing algorithm for task placement cannot find enough space with existing TMs, the Kubernetes Operator can spawn a new TM ``pod'' in the cluster.

%
%
%
%
%
\vspacebeforesection
\section{Evaluation}
\vspaceaftersection
\label{sec:evaluation}


Our evaluation aims to answer the following research questions:
\begin{itemize}
  \item For a given target rate, is \sysname able to converge to efficient configurations in terms of CPU and memory usage?
  \item Does the integration of vertical and horizontal scaling lead to a longer convergence time compared to horizontal (i.e., CPU-only) scaling?
\end{itemize}

We compare \sysname and the \flink DS2 implementation and use the obtained rate (i.e., the capacity of a configuration), the sum of assigned CPU cores, and the sum of allocated memory as metrics.
We use as workload the standard Nexmark Benchmark~\cite{tucker2008nexmark}.
Nexmark simulates an auction system and provides a set of representative queries for the evaluation of batch and stream processing systems.

We use the same six queries as used in the original evaluation of DS2~\cite{kalavri2018three}.
All queries use a single source and a single sink operator.
\textbf{\texttt{Q1}} produces currency conversions using one Map operator.
\textbf{\texttt{Q2}} uses a Filter operator to select bids with a specific identifier.
\textbf{\texttt{Q3}} classifies sales based on location and categories and maps them to specific auctions, using an incremental join operator over the complete stream (i.e., without windowing) and two filter operators.
\textbf{\texttt{Q5}} determines the auctions with the most bids in a period and uses one stateful operator and a group-by-aggregate over a sliding window.
\textbf{\texttt{Q8}} monitors new active users over a period and uses a tumbling-windowed join as a stateful operator.
Finally, \textbf{\texttt{Q11}} monitors user sessions by computing the number of bids each user makes while active.
It uses one operator, a stateful group-by-aggregate, over a session window.


\mypara{Experimental setup}
We alternatively deploy \flink with \sysname or DS2 on a testbed of 7 nodes, each equipped with two 10-core Intel Xeon E5-2630L v4 CPUs, 128~GB of memory, and a 400~GB SSD.
Nodes are connected with 10-Gbps Ethernet.
One node hosts the Kubernetes controller, and another hosts Prometheus and Grafana.
We deploy the Job Manager on another dedicated node.
The four remaining nodes are used to host TMs.

We configure each TM to use 4 CPU cores and 2~GB of memory, shared among 4 TSs.
\flink reserves a portion of the available memory for framework management data and JVM-specific data.
The default amount of managed memory per TS is 158~MB.
With DS2, all tasks get this fixed amount.
With \sysname vertical scaling, a task may receive from 158~MB ($o_i.m^t=0$) to 632~MB ($o_i.m^t=2$, i.e., after two scale-up actions) of managed memory. 

We set elastic scaling trigger parameters to keep the average busyness of operators between 20\% and 80\%.
Through experimental analysis, we identify two thresholds that allow suitable identification of memory pressure for some operator's tasks: A cache hit rate over $\Delta_\theta = 80\%$ and an average state access latency over $\Delta_\tau = 1\text{\emph{ms}}$.
Similarly to DS2, we use short 2-minute decision windows and a 1-minute stabilization period.
If no scaling is triggered at the end of the window, the autoscaler waits for the following full metrics window collection.
Metrics are collected and aggregated with a granularity of 5 seconds.

\begin{figure}[]
    \centering
    \begin{subfigure}[b]{0.48\textwidth}
        \centering
        \includegraphics[scale=0.6]{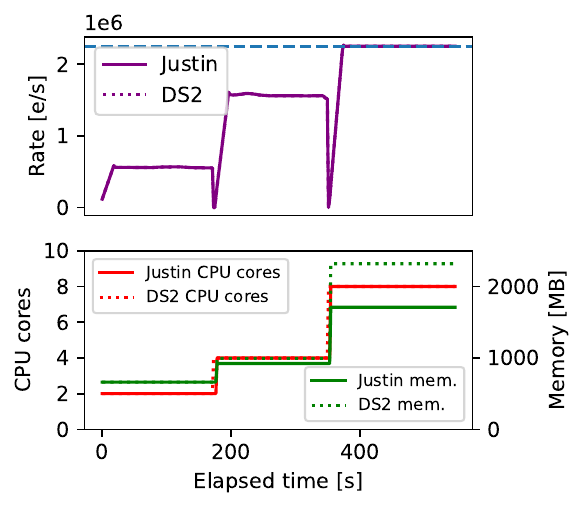}
        \caption{Q1}
        \label{fig:q1}
    \end{subfigure}
    \hfill
    \begin{subfigure}[b]{0.48\textwidth}
        \centering
        \includegraphics[scale=0.6]{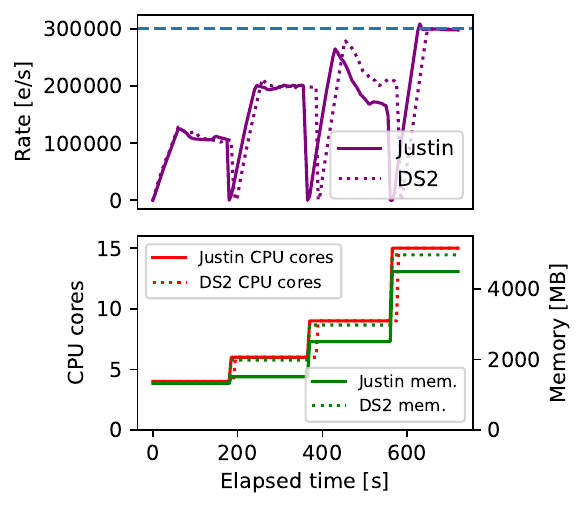}
        \caption{Q3}
        \label{fig:q3}
    \end{subfigure}

    \begin{subfigure}[b]{0.48\textwidth}
        \centering
        \includegraphics[scale=0.6]{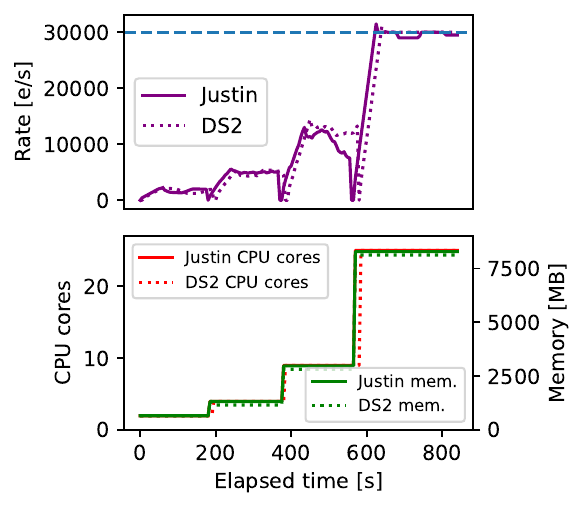}
        \caption{Q5}
        \label{fig:q5}
    \end{subfigure}
    \hfill
    \begin{subfigure}[b]{0.48\textwidth}
        \centering
        \includegraphics[scale=0.6]{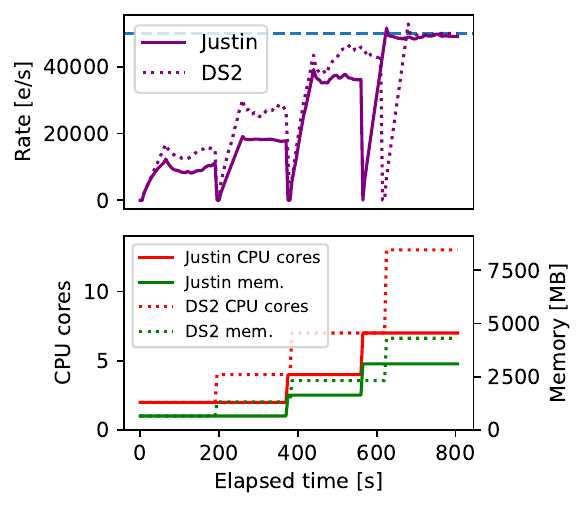}
        \caption{Q11}
        \label{fig:q11}
    \end{subfigure}

    \begin{subfigure}[b]{0.99\textwidth}
        \centering
        \includegraphics[scale=0.6]{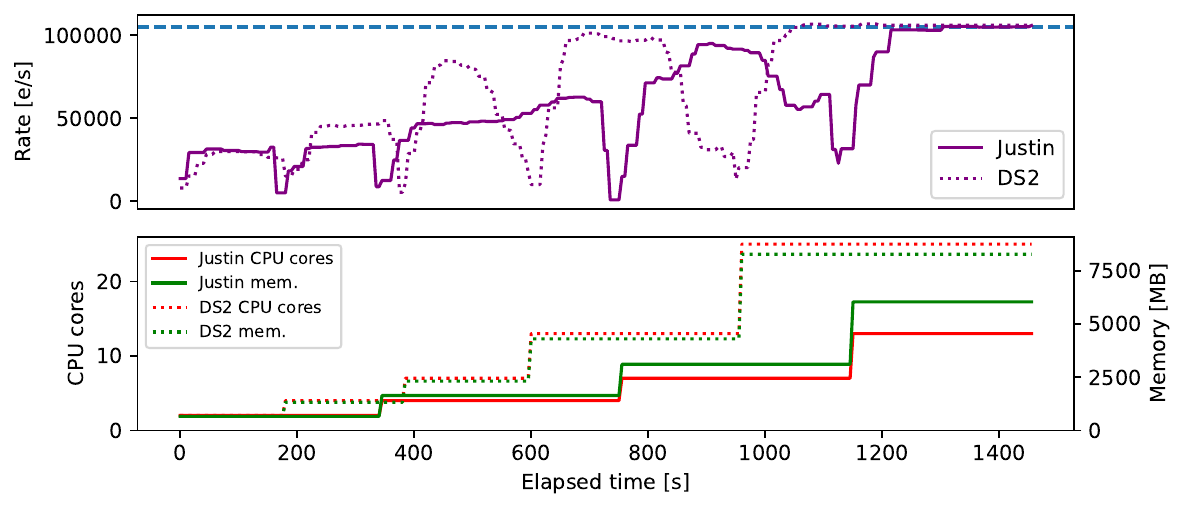}
        \caption{Q8}
        \label{fig:q8}
    \end{subfigure}

    \caption{Elastic scaling performance of \sysname (plain lines) compared to DS2 (dotted lines), achieved capacity (purple), and resulting overall CPU cores (red) and memory consumption (green). The horizontal dashed blue line represents the target rate.}
    \label{fig:nexmark}
\end{figure}

\subsection{Results}
\label{sec:eval:results}

Figure~\ref{fig:nexmark} presents elastic scaling results for \sysname and DS2.
We show the achieved rate (capacity) and the resource allocation for the two auto-scalers as a function of time.
The objective is for sources to reach the target rate indicated by a thin, dashed blue line after a few reconfigurations.
The top plots show the achieved rate, while the bottom plots show the overall CPU and memory consumption (including heap, network, and managed memory).
Note that, as in the DS2 paper~\cite{kalavri2018three}, we exclude sources from the resource count as these are used as stateless workload injectors that would be replaced by lighter-load \flink sources connected to external sources in production.
We include sink operators with a fixed parallelism of one task.
Sinks have a low load across all queries and never are a bottleneck.


\mypara{\texttt{Q1} and \texttt{Q2}: stateless, single-operator queries}
These two queries feature a single stateless operator.
As the results for the two were highly similar, we choose only to show those for \texttt{Q1}.
Figure~\ref{fig:q1} shows that both auto-scalers use two steps to reach a final configuration supporting the 2,250,000 event/s target rate.
Both reach a parallelism of 7 tasks for the operator, but \sysname disables its managed memory.
As a result, the \sysname configuration uses 40\% less memory, from 2,317~MB for DS2 down to 1,379~MB.

\mypara{\texttt{Q3}: two stateless and one stateful operator}
\texttt{Q3} performs an unbounded incremental join, but the size of this operator state converges to a relatively small value ($\sim$8~MB), indicating vertical scaling is likely counterproductive.
The two other operators are stateless.
Figure~\ref{fig:q3} shows that \sysname can free the managed memory of the stateless operators while avoiding unnecessary scale-up for the stateful one, using 10\% less memory than DS2.
Both auto-scalers converge in the same number of steps.
In detail, at the first reconfiguration ($t=1$, 180s), stateless operators get configured as (1; $\bot$) in \sysname and (1; 158) in DS2; after two more scaling steps ($t=3$) the configuration of the stateful operator becomes (12; 158) in both cases.


\mypara{\texttt{Q5}: stateful sliding-window aggregation query}
\texttt{Q5} performs an aggregation over a sliding window, requiring complex access patterns to state in order to re-compute outputs over a new window frequently.
Under the Nexmark workload, the state remains small (i.e., $\sim$10~MB).
As for \texttt{Q3}, vertical scaling would have no impact on the performance of this query.
\sysname chooses only to perform horizontal scaling.
Yet, it saves a small amount of memory by removing the managed memory of the sink operator, as shown by Figure~\ref{fig:q5}.
For both auto-scalers, the final configuration for \texttt{Q5}'s stateful operator is (24; 158).

We discuss \texttt{Q11} before \texttt{Q8} to follow the order of presentation in Figure~\ref{fig:nexmark}.

\mypara{\texttt{Q11}: stateful session window aggregation query}
\texttt{Q11} results in Figure~\ref{fig:q11} illustrate the benefit of replacing a scale-out decision with a scale-up.

We observe a first reconfiguration a little before 200~s for both auto-scalers.
While DS2 increases the parallelism of the primary operator to 3, growing the memory use accordingly, the CPU and memory of \sysname remain equal but for a higher resulting capacity.
This is explained by the joint decisions of \sysname of (1)~stripping the sink operator of its unnecessary managed memory and (2)~allocating the same amount of memory to the primary operator due to a scale-up decision (i.e., increasing its memory level from 0 to 1 while keeping a single task).

The resulting capacity is lower with \sysname but higher when accounted for per expanded core.
Follow-up reconfigurations at $\sim$380~s and $\sim$560~s use scale out for both \sysname and DS2.
However, the better per-task performance of the configuration resulting from the first reconfiguration leads \sysname to require 48\% lower CPU and a 28\% lower memory utilization.
The final configuration matching the target rate is (6; 316) for \sysname and (12; 158) for DS2.
\sysname uses the same number of reconfigurations but converges slightly faster than DS2 with this query.


\mypara{\texttt{Q8}: stateful tumbling window join}
Finally, Figure~\ref{fig:q8} presents results for \texttt{Q8}.
This query is more complex and requires more reconfigurations to achieve the target rate.
As for \texttt{Q11}, the first reconfiguration of DS2 uses a scale-out while \sysname triggers a scale-up of the primary stateful operator.
Interestingly, and in contrast with \texttt{q11}, the scale-up of \sysname seems to have no real benefit, which may lead to thinking one additional round of reconfiguration will be necessary to compensate for a bad decision.
The following reconfigurations show the situation is the opposite.
\sysname performs three scale-out reconfigurations to reach the target rate, while DS2 needs four.
We note that \sysname intermediate configurations take longer to stabilize than with DS2, resulting in a shorter convergence time for the latter despite the additional step.
In terms of resources, however, the advantage is clearly for \sysname, with 48\% less CPU cores and 27\% less memory than DS2 for the same capacity.
The primary operator's final configuration is (12; 316) for \sysname and (24; 158) for DS2.


\mypara{Discussion}
Our experiments with Nexmark show that \sysname exploits hybrid CPU/memory scaling effectively for all queries but one.
It saves memory by stripping unnecessary memory from stateless operators (in primary operators of \texttt{Q1}, \texttt{Q2}, and \texttt{Q3}, and for sink operators of all queries).
It significantly reduces the overall resource consumption for a given target rate in complex stateful queries (\texttt{Q8} and \texttt{Q11}), both in terms of memory as expected and, perhaps more unexpectedly, CPU.
This latter result is due to the higher efficiency of tasks not constrained by state access that, as a result, do not need to reach high parallelism for the same capacity as constrained ones.
\sysname results in fewer or the same number of steps as DS2 and comparable convergence time.
Finally, we observe that for a query that does not really benefit from hybrid CPU/memory scaling (\texttt{Q5}), \sysname does not introduce a penalty over DS2.

\vspacebeforesection
\section{Related Work}
\vspaceaftersection
\label{sec:related}

Elastic scaling for stream processing has gathered significant interest over the last decades.
Representative works include MEAD~\cite{russo2021mead}, hierarchical auto-scaling~\cite{russo2023hierarchical}, DS2~\cite{kalavri2018three}, and more~\cite{schneider2009elastic,gedik2013elastic,barazzutti2014elastic}.
Surveys by R{\"o}ger and Mayer~\cite{roger2019comprehensive} and Cardellini \emph{et al.}~\cite{cardellini2022runtime} provide a comprehensive analysis of this field.
To the best of our knowledge, no elastic scaling method combining dynamic memory allocation with horizontal scaling has been proposed by previous work on the topic.

Resource management in distributed stream processing has also attracted significant interest.
A survey by Liu \emph{et al.}~\cite{liu2020resource} presents an overview of the topic.
Integrating state management with stream processing operators led to advances in programmability and fault tolerance~\cite{carbone2017state,madsen2015dynamic,affetti2017flowdb}.
StreamBed~\cite{rosinosky2024streambed} uses interpolation techniques from guided small-scale runs to build a model and predict resource consumption, including CPU and memory, for large-scale \flink instances.
Recently, Wang \emph{et al.}~\cite{wang2025capsys} proposed CAPSys, a new algorithm for assigning task slots to tasks in \flink that takes into account colocation effects between tasks and competition for resources such as memory and I/O.
CAPSys reduces the unpredictability of round-robin or random task-to-TM assignments.
Adapting its approach to the placement problem in \sysname, where tasks have different memory granularity, is undoubtedly an interesting perspective of our work.

\er{removed a paragraph that is not necessary for dais, but can be reused for a longer version}

Understanding the performance of LSM-based storage~\cite{o1996log,luo2020lsm}, including its relation to available memory, is a complex problem.
Gadget~\cite{asyabi2022new} is a benchmark of backend storage operations in a DSP, targeting \flink and RocksDB.
Tutorials by Sarkar \emph{et al.}~\cite{sarkar2022dissecting,sarkar2023lsm} are a good source of references on optimizations for LSM-based stores and optimization of read operations.

Memory management and allocation is an important topic in non-stream (i.e., batch) processing systems.
MespaConfig~\cite{zong2021mespaconfig} optimizes the memory configuration of \emph{multiple} Spark batch processing jobs co-located on a cluster.
Iorgulescu \emph{et al.}~\cite{iorgulescu2017don} define the memory elasticity principle and show how careful memory allocation below the working set size of batch processing, in-memory applications can yield close performance to larger allocations and implement resulting scheduling policies in Apache Yarn.
The specificities of write-optimized, LSM-based storage in RocksDB prevent from directly applying these techniques.

Finally, we note that the idea of combining horizontal and vertical scaling in a hybrid auto-scaler has been explored in other contexts, such as cloud workloads, e.g., with HoloScale~\cite{millnert2020holoscale}, or for serving model inference~\cite{razavi2024tale}.

\vspacebeforesection
\section{Conclusion}
\vspaceaftersection
\label{sec:conclusion}

\er{To mention:
\begin{itemize}
  \item recap of the contributions
  \item perspectives include:
  \begin{itemize}
    \item take into account skew
    \item allow developers to include memory profiles to their operators
    \item use ML to determine features for operators w.r.t. their best memory and storage configuration
    \item consider the allocation of resources for multiple co-deployed queries in order to mix memory-intensive tasks with cpu-intensive ones that do not require memory and increase the level of consolidation.
  \end{itemize}
\end{itemize}
}

We presented \sysname, a hybrid CPU and memory auto-scaler for \flinklong.
In contrast with DS2, \flink current auto-scaler integrated with its Kubernetes operator, \sysname arbitrates between horizontal scaling and vertical scaling based on metrics from the RocksDB storage layer.
Results on the classical Nexmark benchmark show that \sysname can achieve similar capacity as DS2 when reconfiguring towards a target rate while using fewer resources. 

This work opens several interesting perspectives.
First, and similarly to DS2, \sysname makes implicit assumptions about the absence of skew, i.e., unbalanced key popularities leading to an imbalanced load between the different tasks of an operator.
A solution to this problem is explicitly rebalancing keys between tasks instead of using only a hash function~\cite{nasir2016two,fang2017parallel}.
This does not address the fact that, in some cases, popular keys may be associated with a more significant state.
In this case, heterogeneous memory allocation between tasks of the same operator could be an option, although it would require a significantly more complex auto-scaler.
Second, a complementary path to \sysname reactive auto-scaling approach would be to predict operators' response to memory availability, either by allowing programmers to provide hints or by running them in isolation and modeling their performance~\cite{agnihotri2024zerotune}.
Finally, \sysname and DS2 consider the auto-scaling and resource allocation of queries in isolation.
An auto-scaler that considers the co-placement of query tasks, e.g., running memory-intensive tasks of some queries with stateless tasks of another query, may lead to better consolidation and platform resource use overall.

\er{add link to the trace paper when the paper is accepted (on arxiv if necessary).}

\mypara{Acknowledgments}
This research was funded by the Walloon region (Belgium) through the Win2Wal project ``GEPICIAD'' and by a gift from Eura Nova.
Experiments presented in this paper were carried out using the Grid'5000 testbed, supported by a scientific interest group hosted by Inria and including CNRS, RENATER and several Universities as well as other organizations (see \url{https://www.grid5000.fr}).

%
%
 \bibliographystyle{splncs04}
 \bibliography{references} 

\end{document}